\documentclass[11pt]{article}
\usepackage[paper=a4paper,dvips,top=2.0cm,left=2.2cm,right=2.2cm,bottom=2.6cm,footskip=2.2cm,voffset=0.0cm, marginparwidth=18mm]{geometry}
\usepackage{amsfonts,amsmath,amsthm,amssymb}
\numberwithin{equation}{section} 
\usepackage[svgnames]{xcolor}
\usepackage{fix-cm}
\usepackage{sectsty}
\usepackage{fancyhdr}
\usepackage{lastpage}
\usepackage{graphicx}
\usepackage{url}
\usepackage{enumerate}
\usepackage{paralist}
\usepackage{multicol}
\usepackage{color}  
\usepackage{float}  
\usepackage{comment}
\usepackage{braket}
\usepackage{epsfig}
\usepackage{subfigure}
\usepackage{hyperref}   
\hypersetup{colorlinks=true,linkcolor=beamer@PRD, citecolor=beamer@PRD}
\usepackage{authblk}  
\usepackage{cite}  
\usepackage{todonotes}
\usepackage{ulem} 

\newcommand{\mxout}[1]{\Red{\text{\xout{\ensuremath{#1}}}}} 

\renewcommand\eqref[1]{\textcolor{beamer@PRD}{(}\ref{#1}\textcolor{beamer@PRD}{)}}
\newcommand{\Red}[1]{{#1}}

\definecolor{beamer@PRD}{RGB}{46,48,146}

\begin{document}
\title{\textbf{Probing Planck scale effects on absolute mass limit in neutrino flavor evolution}}
\author{\textbf{Kartik Joshi$^{1}$, Sanjib Dey$^{2,\ast}$ and Satyajit Jena$^{1}$} \\ \small{${}^1$Department of Physical Sciences, Indian Institute of Science Education and Research Mohali, \\ Sector 81, SAS Nagar, Manauli 140306, India\\ ${}^2$Department of Physics, Birla Institute of Technology and Science, Pilani, K K Birla Goa Campus,\\ Zuarinagar, Sancoale, Goa 403726, India \\ $^\ast$E-mail: sanjibd@goa.bits-pilani.ac.in}}
\date{}
\maketitle
\begin{abstract}
This work explores how the generalized uncertainty principle, a theoretical modification of the Heisenberg uncertainty principle inspired by quantum gravity, affects neutrino flavor oscillations. By extending the standard two-flavor neutrino model, we show that the oscillation probability acquires an additional phase term that depends on the {square roots of the individual neutrino masses}, introducing new features beyond the conventional mass-squared differences. To account for the non-Hermitian nature of the resulting dynamics, we employ parity-time ($PT$) symmetric quantum mechanics, which allows for consistent descriptions of systems with {balanced gain and loss mechanisms}. We analyze the feasibility of observing these effects in current and future neutrino experiments, such as DUNE, JUNO, IceCube, ORCA--KM3NeT, MINOS, Daya Bay, Hyper-Kamiokande, and KATRIN, and find that the predicted modifications could fall within the sensitivity of current experiments. Moreover, we propose that analog quantum simulation platforms, such as cold atoms, trapped ions, and photonic systems, offer a promising route to test these predictions under controlled conditions. Our findings suggest that neutrino oscillations may serve as an effective probe of quantum gravity effects, providing a novel connection between fundamental theory and experimental observables.
\end{abstract}
	 
\section{Introduction} \label{Introduction}
\addtolength{\footskip}{-0.2cm} 
 {The unification of quantum mechanics and gravity remains} one of the foremost challenges in modern theoretical physics. A common feature across many quantum gravity candidates, including string theory, loop quantum gravity, and noncommutative geometry, is the prediction of a minimal measurable length scale \cite{KK, CR, Petruzziello}. This leads to modifications of the canonical Heisenberg uncertainty principle, giving rise to the generalized uncertainty principle (GUP) \cite{Boney96}. The GUP introduces corrections to the commutation relations between position and momentum \cite{Kempf_1995, BB, Ali_Das_Vagenas, SD}, which, in turn, can alter the dynamics of quantum systems at high energies or short distances \cite{MG1}. These effects, although typically associated with Planck-scale physics, could manifest as subtle yet measurable deviations in systems exhibiting extreme sensitivity, {notably in phenomena such as neutrino oscillations}.

Neutrinos, once believed to be massless within the Standard Model, have since been shown to possess small but non-zero masses, a discovery confirmed by numerous neutrino oscillation experiments \cite{LW, YF, Nakamura, Blasone, Wang}. Neutrino oscillation arises due to the mismatch between the flavor and mass eigenstates \cite{CG}, with flavor transitions governed by mass-squared differences and mixing angles \cite{CY}. These parameters have been determined with increasing precision in both vacuum and matter scenarios \cite{SP}. However, conventional quantum mechanics does not allow direct access to absolute neutrino masses \cite{JB}, motivating theoretical extensions that could reveal deeper insights.

The GUP modifies the energy-momentum dispersion relation and can therefore affect neutrino propagation phases over long distances. These corrections, though suppressed, may accumulate sufficiently in long-baseline experiments \cite{KJ}, making neutrino oscillations a promising probe of quantum gravitational effects \cite{MS2, Luciano}. Moreover, GUP corrections may lead to mass-dependent phase shifts beyond standard $\Delta m^2$-driven oscillation terms, potentially offering sensitivity to quantities such as $\sqrt{m_1}-\sqrt{m_2}$ rather than just mass-squared differences.

In parallel, another important development is the use of non-Hermitian but $PT$-symmetric operators in quantum mechanics \cite{Bender, AM, CM}. Unlike traditional Hermitian Hamiltonians, $PT$-symmetric systems can admit real energy spectra under certain conditions, even in the presence of gain and loss \cite{Bender_Hook}. These operators have proven effective in modeling open quantum systems, including those involving interactions with external environments \cite{Dey_Raj_Goyal}. Notably, they offer mathematical consistency while extending the descriptive power of quantum theory. In neutrino physics, where weak interactions violate parity ($P$) and time-reversal ($T$) symmetries individually, but not necessarily in combination ($CPT$ symmetry is preserved), $PT$-symmetric formulations provide a natural setting to study flavor evolution and possible quantum gravity effects \cite{TO}.

In this work, we explore the combined impact of GUP and $PT$ symmetry on neutrino oscillations by introducing $PT$-symmetric deformations into the Schrödinger evolution of a two-flavor neutrino system. Starting from GUP-modified canonical commutation relations, we derive the resulting corrections to the neutrino oscillation probability and demonstrate that the modified phase includes a term proportional to $\sqrt{m_1}-\sqrt{m_2}$, in addition to the standard $\Delta m^2/2E$ term. This leads to a new class of oscillation behavior that can, in principle, be constrained or detected in current and upcoming neutrino experiments such as DUNE \cite{YF}, JUNO \cite{Abi2020, JUNO2022}, NO$\nu$A, Super-Kamiokande \cite{Abe2018}, Hyper-Kamiokande \cite{hyper, hyper1,hyper2}, and KATRIN.

Furthermore, we analyze the feasibility of detecting these GUP-induced corrections using realistic experimental baselines and energy ranges. To address the challenges posed by ultra-small phase differences, we also propose analog quantum simulation frameworks as complementary platforms. Systems such as cold atoms, ion traps, and photonic lattices, employing nonclassical states and interferometric techniques, can mimic neutrino dynamics with tunable parameters, achieving phase resolution far beyond conventional detectors \cite{Pezze2018,Duttatreya}. Such approaches open exciting new avenues to simulate and test Planck-scale physics in the laboratory \cite{Pikovski,Dey_NPB, Khodadi}.

Taken together, this study positions neutrino oscillations not only as a powerful tool for understanding the neutrino sector itself but also as a sensitive probe of fundamental quantum structure, with potential implications for both quantum gravity phenomenology and non-Hermitian quantum theory.

Our article is organized as follows. In Sec.\,\ref{sec2}, we introduce the GUP model and provide solutions to the GUP-modified Schr\"odinger equation. Sec.\,\ref{sec3} is equipped with the GUP-modified neutrino oscillation framework, where we calculate the GUP-modified transition probability. In Sec.\,\ref{sec4}, we discuss the experimental feasibility for detecting the GUP-modified oscillation phase. Sec.\,\ref{sec5} investigates potential constraints on the existing and future experiments, and finally, in Sec\,\ref{sec7}, we conclude our results and provide an outlook of our study.
\section{Generalized uncertainty principle}\label{sec2}
\lhead{Planck-scale effects on neutrino mass and flavor evolution}
\chead{}
\rhead{}
\addtolength{\voffset}{0.8cm} 
\addtolength{\footskip}{-0.7cm} 
GUP, an extension of Heisenberg’s uncertainty relations, has been widely studied for its role in predicting minimal lengths and momenta \cite{Kempf_1995, BB, Ali_Das_Vagenas, SD}. Deformed canonical variables arising from the GUP are shown to correspond to noncommutative spacetimes, with creation and annihilation operators following $q$-deformed oscillator algebras \cite{SD}. Here, we use a one-dimensional noncommutative space, obtained as a simplified component of the three-dimensional model in \cite{SD}
\begin{equation}\label{GUP1}
    [X,P]=i\hbar (1+\beta P^2),
\end{equation}
where $\beta=\tilde{\beta}/(\hbar m \omega)$, with $\tilde{\beta}$ being dimensionless. Several representations of \eqref{GUP1} have been reported in \cite{HK}, among which we choose the following
\begin{equation}\label{Rep}
     X=x, \qquad  P=\frac{1}{\sqrt{\beta}}\tan\left(\sqrt{\beta}p\right),
\end{equation}
where $x$ and $p$ represent the standard canonical variables satisfying $[x,p]=i\hbar$. 
\subsection{Solution of modified Schr\"odinger equation}
Considering the representation \eqref{Rep}, the Hamiltonian of a free particle is modified 
\begin{equation}\label{GUPHam}
    H^{GUP}=\frac{P^2}{2m} = \frac{1}{2 m \beta}\left(\sqrt{\beta}p+ \frac{1}{3}\beta\sqrt{\beta}p^3+...\right)^2,
\end{equation}
thus yielding the non-relativistic time-independent Schr\"odinger equation, up to the first order of $\beta$, turns out to be \cite{GH}
\begin{equation}\label{Modified_SE}
        \frac{2\beta\hbar^4}{3} \psi{''''} -\hbar^2\psi^{''} -2mE\psi=0.
\end{equation}
The Hamiltonian \eqref{GUPHam} presents an effective low-momentum expansion of the GUP-modified Hamiltonian.  The solution of \eqref{Modified_SE} is given by \cite{CR}
\begin{equation}\label{Modified_WF}
    \psi(x)= c_{1}e^{ik_{1}x}+c_{2}e^{-ik_{1}x}+c_{3}e^{k_{2}x}+c_{4}e^{-k_{2}x},
\end{equation}
with $k_{1}^2=\frac{3}{4\beta \hbar^2}\left(\alpha-1\right)$ and $k_{2}^2=\frac{3}{4\beta \hbar^2}\left(\alpha+1\right)$, where $\alpha = \sqrt{1+\frac{16}{3}\beta mE}$. The wave function \eqref{Modified_WF} remains normalizable and satisfies all basic properties of a well-behaved physical wave function \cite{TO}. For further details on this, refer to the Appendix \ref{A1}. For finite boundary conditions, all constants $c_1, c_2, c_3, c_4$ can be non-zero. However, under infinite boundary conditions, certain terms must vanish to avoid non-normalizable solutions. For instance, setting $c_3$ and $c_4$ to zero leads to solutions where $k_1 = \sqrt{2mE}$.

\section{GUP modified neutrino oscillation}\label{sec3}
Let \( \ket{\nu_{1}} \) and \( \ket{\nu_{2}} \) be two neutrino mass eigenstates with masses \( m_{1} \) and \( m_{2} \), respectively, and \( \ket{\nu_{x}} \) and \( \ket{\nu_{y}} \) be the flavor bases \cite{LW}. Then, the flavor states can be written in terms of the mass eigenbasis as
 \begin{eqnarray} \label{Eigen1}
     \ket{\nu_{x}} &=& \cos\theta \ket{\nu_{1}} + \sin \theta \ket{\nu_{2}}, \\
     \ket{\nu_{y}} &=& -\sin\theta \ket{\nu_{1}} + \cos \theta \ket{\nu_{2}}, \nonumber
 \end{eqnarray}
thereby yielding the mass eigenstates in terms of the flavor basis can be written as
\begin{equation}\label{Eigen2}
\begin{pmatrix}
\ket{\nu_{1}} \\
\ket{\nu_{2}}
\end{pmatrix} = 
\begin{pmatrix}
\cos \theta & -\sin \theta \\
\sin \theta & \cos \theta
\end{pmatrix} 
\begin{pmatrix}
\ket{\nu_{x}} \\
\ket{\nu_{y}}
\end{pmatrix}.
\end{equation}
Let a flavor $x$ neutrino \( \ket{\nu_{x}(0)} \) be produced at $t=0$, and its time evolution is governed by the modified Schr\"{o}dinger equation \eqref{Modified_SE}. Accordingly, the time-evolved mass eigenstates are given by
\begin{equation}\label{Eigen3}
    \begin{pmatrix}
    \ket{\nu_{1}(t)} \\
    \ket{\nu_{2}(t)}
    \end{pmatrix}= 
    \begin{pmatrix}
    \gamma_{1} & 0 \\
    0 & \gamma_{2}
    \end{pmatrix}
    \begin{pmatrix}
    \ket{\nu_{1}(0)} \\
    \ket{\nu_{2}(0)}
    \end{pmatrix},
\end{equation}
where 
\begin{equation}\label{gamma_j}
\gamma_{j}= \exp\left[-iE_{j}t + i\sqrt{\frac{3}{4\beta \hbar^2}\left(\alpha_j-1\right)}x\right], \quad \text{with } \alpha_j = \sqrt{1 + \frac{16}{3}\beta m_j E_j}, \quad j = (1, 2).
\end{equation}
Thus, the time-dependent flavour states \eqref{Eigen1} with the help of \eqref{Eigen3} become
\begin{equation}
        \begin{pmatrix}
\ket{\nu_{x}(t)} \\
\ket{\nu_{y}(t)}
\end{pmatrix}   =
\begin{pmatrix}
\cos \theta & \sin \theta \\
-\sin \theta & \cos \theta
\end{pmatrix} 
   \begin{pmatrix}
\gamma_{1} & 0 \\
0 & \gamma_{2}
\end{pmatrix} 
\begin{pmatrix}
\ket{\nu_{1}(0)} \\
\ket{\nu_{2}(0)}
\end{pmatrix}.
\end{equation}
Subsequently, using \eqref{Eigen2}, the time-evolved flavour states in the flavour basis are modified to
\begin{equation}
        \begin{pmatrix}
\ket{\nu_{x}(t)} \\
\ket{\nu_{y}(t)}
\end{pmatrix}   =
\begin{pmatrix}
\cos \theta & \sin \theta \\
-\sin \theta & \cos \theta
\end{pmatrix} 
   \begin{pmatrix}
\gamma_{1} & 0 \\
0 & \gamma_{2}
\end{pmatrix} 
\begin{pmatrix}
\cos \theta & -\sin \theta \\
\sin \theta & \cos \theta
\end{pmatrix} 
\begin{pmatrix}
\ket{\nu_{x}(0)} \\
\ket{\nu_{y}(0)}
\end{pmatrix},
\end{equation}
which, when simplified, takes the form
\begin{equation}
        \begin{pmatrix}
\ket{\nu_{x}(t)} \\
\ket{\nu_{y}(t)}
\end{pmatrix}   
 \begin{pmatrix}
\cos^2 \theta \gamma_{1}+ \sin^2\theta \gamma_{2} & -\sin \theta \cos \theta(\gamma_{1}-\gamma_{2})\\
-\sin \theta \cos \theta(\gamma_{1}-\gamma_{2}) & \cos^2 \theta \gamma_{2}+ \sin^2\theta \gamma_{1}
\end{pmatrix} 
\begin{pmatrix}
\ket{\nu_{x}(0)} \\
\ket{\nu_{y}(0)}
\end{pmatrix}.
\end{equation}
Therefore, the probability of a neutrino changing its flavor from $x$ to $y$ in time $t$ is
\begin{equation} \label{ProbN}
P(\nu_x \to \nu_y) = \left|\langle \nu_y(t) | \nu_x(0) \rangle\right|^2 = \sin^2 \theta \cos^2 \theta \left|\gamma_1 - \gamma_2\right|^2.
\end{equation}
Assuming $\beta m_j E_j \ll 1$, we can modify $\alpha_j$ in \eqref{gamma_j} as $\alpha_j \approx 1 + \frac{8}{3} \beta m_j E_j$, so that
\begin{equation}\label{gamma_j_1}
\gamma_j=exp\left[-iE_j t + i\sqrt{\frac{2m_jE_j}{\hbar^2}}\right].
\end{equation}
Furthermore, using the relativistic dispersion relation for neutrinos $E_j \approx p + \frac{m_j^2}{2p}$ with $p>>m_j$, we rewrite \eqref{gamma_j_1} as
\begin{align}
\gamma_j &\approx \exp\left[-i\left(p + \frac{m_j^2}{2p}\right) t + i \frac{\sqrt{2 m_j p}}{\hbar} \left(1 + \frac{m_j^2}{4p^2}\right) x\right]=e^{iA_j}, \quad j=(1,2),
\end{align}
where $A_j=-\left(p + \frac{m_j^2}{2p}\right) t +  \frac{\sqrt{2 m_j p}}{\hbar} \left(1 + \frac{m_j^2}{4p^2}\right) x$. Therefore
\begin{alignat}{1}
|\gamma_1-\gamma_2|^2 &= \left\vert e^{iA_2}\left[e^{i(A_1-A_2)}-1\right] \right\vert^2 = 4\sin^2 \left(\frac{A_1-A_2}{2}\right) \nonumber \\
&= 4 \sin^2\left[ \frac{1}{2} \left\{ \frac{m_2^2-m_1^2}{2p}t + \frac{x \sqrt{2p}}{\hbar} \left( \sqrt{m_1} - \sqrt{m_2} + \frac{1}{4p^2}(m_1^2 \sqrt{m_1} - m_2^2 \sqrt{m_2}) \right) \right\} \right] \label{g1g2}
\end{alignat}
where we have used the trigonometric identity $|e^{i\theta} - 1|^2 = 4 \sin^2(\theta/2)$. Given $p^2 \gg m_{i}^2$, equation \eqref{g1g2} is simplified as
\begin{equation}
|\gamma_1-\gamma_2|^2 = 4 \sin^2\left[ \frac{1}{2} \left\{ \frac{m_2^2-m_1^2}{2p}t + \frac{x \sqrt{2p}}{\hbar} \left( \sqrt{m_1} - \sqrt{m_2}  \right) \right\} \right]. 
\end{equation}
Using natural units, i.e., $\hbar = 1, t=x\approx L$, and $p \approx E$, we obtain 
\begin{equation}
|\gamma_1-\gamma_2|^2 = 4 \sin^2\left[ \frac{1}{2} \left\{ \frac{m_2^2-m_1^2}{2E}L + L \sqrt{2E} \left( \sqrt{m_1} - \sqrt{m_2}  \right) \right\} \right],
\end{equation}
where $L$ is the distance travelled by neutrino flavors. Thus, the GUP modified probability of flavor change \eqref{ProbN} is given by 
\begin{equation}
P(\nu_x \to \nu_y) = \sin^2{2\theta} \sin^2\left[ \frac{L}{2} \left\{ \frac{m_2^2-m_1^2}{2E} + \sqrt{2E} \left( \sqrt{m_1} - \sqrt{m_2}  \right) \right\} \right].
\end{equation}
Here, the second term within the curly brackets represents the GUP modification to the oscillation probability, i.e., the GUP-modified oscillation phase is 
\begin{equation}\label{ModPhase}
\phi \sim \frac{\Delta m^2}{2E} L + L\sqrt{2E} \left( \sqrt{m_1} - \sqrt{m_2}  \right),
\end{equation}
where $\Delta m^2=m_2^2-m_1^2$ represents the mass-squared difference.
\section{Feasibility analysis of the GUP modified phase detection}\label{sec4}
Detecting extremely small phase shifts (on the order of $\mxout{\sim} 10^{-8}$ radians) and probing sub-Planckian neutrino mass differences pose significant challenges in long-baseline neutrino experiments. To address this, we propose an analog quantum simulation framework with three main goals: (i) to simulate the evolution of neutrino phases modified by the GUP within a controllable quantum system, (ii) to reproduce the modified oscillation probabilities incorporating mass-dependent phase terms, and (iii) to assess the sensitivity and practical feasibility of detecting GUP-induced effects.

Current neutrino experiments aiming to measure absolute neutrino masses typically have phase resolution ($\Delta\phi$) capabilities in the range of 0.1 and 0.001 eV \cite{YG}. For atmospheric neutrinos \cite{TK,Abe2018}, the typical ranges are energy $E \approx$ 1-10 GeV and baseline distance $L \approx 10^{4}-10^{5}$ km. Using representative values $E=1 \text{ GeV}$ and $L=10^{-4}$ km in the GUP-modified phase expression from \eqref{ModPhase}, we estimate the minimum measurable difference in the {square roots of the individual neutrino masses} as
\begin{equation}
\sqrt{m_1} - \sqrt{m_2} \gtrsim \frac{\Delta\phi}{L \sqrt{E}} \approx \frac{10^{-3}}{10^{4} \text{km}\sqrt{10^{9} \text{eV}}} \approx 6.25 \times 10^{-22} \text{eV}^{1/2},
\end{equation}
where 1 km $\approx 5.09 \times 10^{9} \text{eV}^{-1}$. It indicates that the GUP-corrected expression, when applied with realistic baselines and energies, which lies within the sensitivity reach of existing detectors, suggesting that existing setups may already be capable of detecting such mass differences, This result implies that any experiment aiming to probe a GUP-induced correction involving the term $\sqrt{m_2}-\sqrt{m_1}$ would require a sensitivity better than $10^{-21}~\text{eV}^{1/2}$—a remarkably small threshold. Only high-statistics, long-baseline, or atmospheric neutrino experiments such as \textsc{DUNE} and \textsc{ORCA} \cite{orca2019, orca2022, orca2023}, or future high-precision interferometric setups, may be capable of approaching this level of sensitivity.

Although current neutrino beams are constrained by detector resolution and other experimental limitations, analog quantum systems offer a promising platform to simulate GUP-induced modifications in neutrino oscillations. These simulations can play a crucial role in placing theoretical bounds on GUP parameters and guiding the design of next-generation neutrino experiments. Furthermore, long-baseline experiments such as T2K, NO$\nu$A, and DUNE operate at precisely controlled energy and baseline configurations, making them well-suited to test for deviations from standard oscillation predictions, including GUP-related effects involving terms like $\sqrt{m_1} - \sqrt{m_2}$ \cite{Abi2020}. Reactor-based experiments like Daya Bay and JUNO, with their precise measurements at MeV energies and short baselines, are particularly effective for probing new physics \cite{JUNO2022}. Atmospheric neutrino detectors like Hyper-Kamiokande \cite{hyper,hyper1,hyper2} and IceCube, which cover a wide range of energies and baselines, provide additional avenues to study energy-dependent GUP-induced corrections.

In the standard two-flavor neutrino oscillation, the survival probability of a neutrino flavor $\nu_\alpha$ after traveling a distance $L$ is given by
\begin{equation}
P_{\alpha \to \alpha}(L) = 1 - \sin^2(2\theta)\, \sin^2\left(\frac{\Delta m^2\,L}{4E}\right),
\end{equation}
where $\Delta m^2$ is the mass-squared difference, $E$ is the neutrino energy, and $\theta$ is the mixing angle. When modifications due to the GUP are taken into account, the oscillation phase receives an additional contribution that depends on the square roots of the neutrino masses. Specifically, the modified phase becomes
\begin{equation}
\Phi = \frac{\Delta m^2\,L}{4E} + \gamma\,L\,\left(\sqrt{m_2} - \sqrt{m_1}\right)\, \sqrt{2E},
\end{equation}
where $\gamma$ is a phenomenological parameter that encapsulates the effects of the GUP. Consequently, the survival probability is altered to
\begin{equation}
P_{\alpha \to \alpha}^{\text{GUP}}(L) = 1 - \sin^2(2\theta)\, \sin^2\left[\frac{\Delta m^2\,L}{4E} + \gamma L\left(\sqrt{m_2} - \sqrt{m_1}\right)\sqrt{2E} \right].
\end{equation}
This additional phase term can lead to measurable distortions in the disappearance spectrum, particularly in high-precision experiments such as Daya Bay and MINOS, which are sensitive to small deviations in the survival probability across a broad range of energies and baselines.

\subsection{Numerical estimation of GUP phase shift}
To assess the detectability of the GUP-induced correction to the neutrino oscillation phase, we provide an order-of-magnitude estimate of the additional phase shift. In the framework where GUP introduces a correction proportional to the square roots of neutrino masses, the modified oscillation phase takes the form
\begin{equation}
\delta \phi_{\text{GUP}} = \gamma L \Delta\sqrt{m} \sqrt{2E},
\end{equation}
where $\gamma$ is the GUP deformation parameter, $L$ is the baseline length, $E$ is the neutrino energy, and $\Delta\sqrt{m}=\sqrt{m_2}-\sqrt{m_1}$ represents the difference in mass roots between neutrino eigenstates.
Assuming representative values such as $L = 1000$ km (MINOS baseline), $E = 1$ GeV, $\sqrt{m_2} - \sqrt{m_1} \sim 10^{-21}$ eV$^{1/2}$, and a target sensitivity of $\delta \phi_{\text{GUP}} \sim 10^{-3}$ rad, we estimate
\begin{eqnarray}
&& \gamma_{\text{GUP}} \sim 1.0 \times 10^{-9}.
\end{eqnarray}
This value is extremely small, suggesting that only future high-precision experiments, capable of resolving oscillation phase shifts down to $\sim 10^{-4}$ radians, could probe such effects. Depending on experimental configurations and underlying neutrino mass assumptions, sensitivity to $\gamma$ in the range $10^{-6} $ to $ 10^{-9}$ may become attainable.

We also revisit an earlier GUP phase estimate in the form 
\begin{equation}
\Delta \phi_{\text{GUP}} = \frac{\gamma p^3 L}{2} \sim 10^{-3}
\end{equation}
where $p\approx E$ in natural units. For a baseline $L = 10^4$ km and a typical neutrino momentum $p = 1$ GeV, requiring a detectable phase shift of $\Delta\phi_{\text{GUP}} \sim 10^{-3}$ leads to $\gamma \sim 2 \times 10^{-34}$. This value is consistent with independent estimates found in the literature and confirms the internal consistency of our framework.

We further express this result in terms of the $\sqrt{m}$-dependent formulation discussed earlier, which arises in alternate GUP models where phase corrections depend on differences in the square roots of mass eigenvalues. The general structure of such corrections, along with their associated coefficients, has been studied in different physical systems, and our formulation aligns with these established approaches \cite{SD}.

MINOS and Daya Bay, operating in accelerator-based and reactor-based regimes respectively, provide high-statistics measurements of disappearance probabilities. This makes them particularly sensitive to small deviations in survival phases, offering complementary constraints on the GUP parameter $\gamma$, alongside long-baseline appearance experiments such as DUNE or T2K.

To ensure consistency with experimental data, we now constrain $\gamma$ such that GUP-induced corrections remain within the perturbative regime. An error-band analysis, illustrating the range of physically allowed values, has been included in Section 5. Additionally, we have now explicitly related the deformation parameter $\gamma$ to the underlying GUP parameter $\beta$ via the identification $\gamma = \beta \hbar^2$ (in natural units). This restores the missing connection and enables experimental bounds on $\gamma$ to be translated into corresponding constraints on $\beta$.

\section{Potential constraints on $\sqrt{m_1} - \sqrt{m_2}$ from existing and future experiments} \label{sec5}
The sensitivity of present and upcoming neutrino experiments to oscillation phases and mass parameters enables the possibility of placing constraints on GUP-induced corrections, particularly the term involving $\sqrt{m_1} - \sqrt{m_2}$ in the modified oscillation probability. In contrast to traditional oscillation experiments that infer mass differences indirectly through mass-squared differences, some precision experiments aim to directly measure the absolute neutrino mass \cite{Aker2019}.

One of the most prominent examples is the KATRIN (Karlsruhe Tritium Neutrino) experiment, which measures the endpoint of the electron energy spectrum in tritium beta decay to infer the mass of the electron antineutrino. KATRIN's primary goal is to improve the upper limit on the electron antineutrino mass to below 0.3 eV/c$^2$ (at $90\%$ confidence level) or detect a non-zero mass if it exceeds 0.35 eV/c$^2$. Currently, it has already set a leading upper bound of 0.45 eV (at 90$\%$ confidence level).

Another key effort is Project 8 \cite{AA}, which is developing a novel detection method called Cyclotron Radiation Emission Spectroscopy (CRES) to measure the energy of electrons from tritium decay. Project 8 aims to achieve sensitivity as low as 40 meV/c$^2$ to the effective neutrino mass, which is a weighted combination of the neutrino mass eigenstates. These direct detection methods are especially valuable for testing the GUP-modified neutrino oscillation framework, where the phase correction includes a dependence on the square roots of the masses \cite{AA}. By comparing precise measurements of the absolute neutrino mass with oscillation data, one can potentially identify or constrain deviations predicted by the GUP model. A summary of relevant experiments and their sensitivity is given in Table \ref{tab:neutrino_experiments_gup_compact}. These experiments are crucial because they measure both standard oscillation parameters and, in some cases, the absolute mass scale, providing an independent cross-validation for models that predict GUP-induced modifications. The precision with which these mass-related quantities are now being measured allows experimental validation or rejection of such theoretical corrections.
\begin{table*}[h!]
\centering
\small 
\begin{tabular}{|p{1.9cm}|p{6cm}|p{3cm}|l|}
\hline
\textbf{Experiment name} & \textbf{Primary goal} & \textbf{Sensitivity to $\Delta m^2$} & \textbf{Limit on $\Delta$m (eV)} \\
\hline
T2K  & Precision on oscillation parameters & $\Delta m^2 \sim 10^{-3}$ eV$^2$ & - \\
\hline
NO$\nu$A  & Mass hierarchy, CP violation & $\Delta m^2 \sim 10^{-3}$ eV$^2$ & - \\
\hline
Hyper-Kamiokande  & Atmospheric and solar neutrinos & $\Delta m^2 \sim 10^{-3}, 10^{-4}$ eV$^2$ & - \\
\hline
IceCube/ DeepCore  & Atmospheric neutrinos & $\Delta m^2 \sim 10^{-3}$ eV$^2$ & - \\
\hline
ORCA--KM3NeT & Mass hierarchy and atmospheric neutrinos in the multi-GeV range & $\Delta m^2 \sim 10^{-3}$ eV$^2$ & -\\
\hline
KATRIN  & Absolute mass of electron antineutrino & - & 0.45 \\
\hline
Project 8  & Electron neutrino mass via CRES & - & 0.04-meV($\sqrt{m_1}-\sqrt{m_2}$ ) \\
\hline
\end{tabular}

\label{tab:current_experiments}

    \caption{\small{Neutrino experiments and their relevance to parameters in the GUP-modified probability. The potential sensitivity to $\sqrt{m_1} - \sqrt{m_2}$ arises from the precision with which these experiments measure standard oscillation parameters and absolute neutrino masses, enabling meaningful comparisons with predictions from GUP-modified models.}}
    \label{tab:neutrino_experiments_gup_compact}
\end{table*}

The introduction of the GUP leads to a modified energy-momentum relation, which in turn alters the oscillation phase and directly impacts the survival and transition probabilities observable in neutrino detection experiments. In standard quantum mechanics, the flavor oscillation probability between neutrino species is given by 
\begin{equation}
P_{\nu_\alpha \rightarrow \nu_\beta}(L) = \delta_{\alpha\beta} - 4 \sum_{j>i} U_{\alpha i} U_{\beta i} U_{\alpha j} U_{\beta j} \sin^2\left(\frac{\Delta m_{ji}^2 L}{4E}\right),
\end{equation}
where $U$ is the PMNS matrix, $\Delta m_{ji}^2$ is the mass-squared difference, $L$ is the baseline, and $E$ is the neutrino energy. Under the GUP framework, the energy-momentum relation is modified. For the simplest GUP,
\begin{equation}
E^2 = p^2 (1 + \gamma p^2) + m^2 \Rightarrow E_i \approx p + \frac{m_i^2}{2p} + \gamma \frac{p^3}{2},
\end{equation}
where $\gamma$ is the GUP-induced deformation parameter. This leads to a modified oscillation phase of the form
\begin{equation}
\Delta\phi_{ij}^{GUP} = \frac{(\Delta m_{ij}^2 + \gamma p^4) L}{2E} \Rightarrow P^{GUP}_{\nu_\alpha \rightarrow \nu_\beta} \sim \sin^2\left(\frac{\Delta m_{ij}^2 L}{4E} + \frac{\gamma p^4 L}{4E}\right)
\end{equation}
indicating that the oscillation probabilities measured in experiments could exhibit measurable shifts due to quantum gravity-induced effects.

The parameter $\gamma$ characterizes the scale at which GUP-induced quantum gravity effects become significant. Since its precise value is model-dependent and currently unknown, we present results in a range of plausible values to illustrate potential signatures \cite{gup2024phen}. Typical range of $\gamma$:
\begin{equation}
\gamma \sim \gamma_0 / M_{Pl}^2 \text{ with } \gamma_0 \in [10^{-1}, 10^{-5}] \Rightarrow \gamma \in [10^{-38}, 10^{-34}] ~\text{eV}^{-2} .
\end{equation}
Although this range lies well below current experimental sensitivity, it serves as a useful benchmark for illustrating possible Planck-scale signatures in neutrino oscillation data. 

While the analysis in this work primarily uses the two-flavor approximation for analytical clarity, the GUP-induced correction can be naturally extended to the full three-flavor formalism. The standard three-flavor Hamiltonian \cite{nonherm4} in the flavor basis is
\begin{equation}
H = \frac{1}{2E} U \begin{pmatrix} 0 & 0 & 0 \\ 0 & \Delta m_{21}^2 & 0 \\ 0 & 0 & \Delta m_{31}^2 \end{pmatrix} U^\dagger ,
\end{equation}
which, in the presence of GUP-induced effects, becomes
\begin{equation}
H^{GUP}_{ij} = H_{ij} + \delta_{ij} \frac{\gamma p^3}{2}.
\end{equation}
This leads to a momentum-dependent diagonal shift in the mass eigenvalues. The time evolution of neutrino flavor states is then governed by
\begin{equation}
i \frac{d}{dt} |\nu(t)\rangle = H^{GUP} |\nu(t)\rangle .
\end{equation}
These corrections can be treated perturbatively, and the formalism may be extended further to incorporate the complex CP phase in the PMNS matrix, which could unveil novel GUP-induced signatures in CP-violating observables.

Although neutrinos typically maintain coherence over distances far exceeding Earth's diameter—typically $10^{3}-10^{5}$ km depending on energy and mass-squared differences- the non-Hermitian structure introduced in our framework does not stem from environmental decoherence. Instead, it emerges intrinsically from the GUP-induced modifications to the quantum Hamiltonian. Specifically, the GUP-altered Hamiltonian modifies the oscillation phase and energy levels, and can induce an imaginary component through higher-order momentum corrections in specific GUP models.

In standard quantum mechanics, the Hermiticity of the Hamiltonian ensures unitary time evolution. However, in the presence of GUP, the modified energy-momentum relation introduces nonlinear and potentially imaginary terms at high momenta. This can result in an effective Hamiltonian of the form
\begin{equation}
H^{\text{GUP}} = H_0 + \delta H = \frac{1}{2E} U \begin{pmatrix} 0 & 0 \\ 0 & \Delta m^2 \end{pmatrix} U^\dagger + i \epsilon(p, \gamma),
\end{equation}
where $i \epsilon(p, \gamma)$ denotes the GUP-induced non-Hermitian corrections at momentum scale $p$ and deformation parameter $\gamma$. Unlike environmental decoherence (e.g., from matter interactions or finite detector resolution), this term arises intrinsically from the modified quantum structure. In our formalism, the effective Hamiltonian can be written in the two-flavor basis as
\begin{equation}
H^{\text{eff}} = \frac{\Delta m^2}{4E} \begin{pmatrix} -\cos 2\theta & \sin 2\theta \\ \sin 2\theta & \cos 2\theta \end{pmatrix} + \gamma p^2 f(p),
\end{equation}
where $f(p)$ is a model-dependent function. Depending on the choice of basis and the form of $f(p)$, this correction may appear in diagonal or off-diagonal elements and can lead to damping in the oscillation probability:
\begin{equation}
P_{\nu_\alpha \to \nu_\beta}(L) \approx \sin^2(2\theta) \exp(-\Gamma L) \sin^2\left(\frac{\Delta m^2 L}{4E} + \gamma p^3 L\right),
\end{equation}
with $\Gamma$ being a phenomenological parameter associated with the imaginary part of the GUP Hamiltonian.

Thus, the GUP-induced non-unitarity explored here is fundamentally different from standard decoherence effects and remains relevant even in vacuum propagation. These intrinsic quantum gravity effects may be probed in high-precision neutrino oscillation experiments. Experiments such as MINOS and Daya Bay are also relevant in exploring potential GUP-induced modifications, given their precision and energy coverage overlapping with the sensitive regions of our model.

To assess the experimental feasibility of detecting GUP-induced effects, we provide sensitivity estimates for the deformation parameter $\gamma$ appearing in the modified oscillation phase. Specifically, the additional GUP-induced contribution to the phase is given by 
\begin{equation}
\Delta\phi_{\text{GUP}} = \frac{\gamma p^3 L}{2},
\end{equation}
where $p$ is the neutrino momentum and $L$ is the baseline. Requiring this correction to induce a minimum observable phase shift, say $\Delta\phi_{\text{GUP}} \gtrsim 10^{-3}$, allows us to estimate the minimum detectable $\gamma$. for different experimental setups.

\begin{table}[ht]
\centering
\renewcommand{\arraystretch}{0.5}
\setlength{\tabcolsep}{4pt}
\begin{tabular}{|l|p{1.6cm}|p{1.6cm}|c|p{6.8cm}|}
\hline
\textbf{Experiment} & \textbf{Baseline $L$(km)} & \textbf{Energy $E$(GeV)} & \textbf{Sensitivity to $\gamma$} & \textbf{Possible Modified Phase Detection} \\
\hline
DUNE       & 1300    & 3.5     & $\gtrsim 7.44 \times 10^{-14}$ & Strong probe via long-baseline $\nu_\mu \rightarrow \nu_e$ channels. \\
\hline
JUNO       & 53      & 0.005   & $\gtrsim 4.83 \times 10^{-11}$ & Limited by low  energy and short baseline. \\
\hline
ORCA       & 5000    & 15      & $\gtrsim 9.34 \times 10^{-15}$ & Excellent match for  high-energy GUP phase sensitivity. \\
\hline
MINOS      & 735     & 5.5     & $\gtrsim 1.05 \times 10^{-13}$ & Archival $\nu_\mu$  survival data useful for testing. \\
\hline
Daya Bay   & 1.2     & 0.004   & $\gtrsim 2.38 \times 10^{-9}$  & Not sensitive due to very low energy and short baseline. \\
\hline
\end{tabular}
\caption{Sensitivity of selected neutrino experiments to the GUP parameter $\gamma$, assuming a minimum detectable phase shift of $10^{-3}$.}
\label{tab:gamma_sensitivity}
\end{table}
The table \ref{tab:gamma_sensitivity} connects the theoretical prediction of a GUP-modified oscillation phase to realistic experimental energy-baseline configurations. While some experiments are more suited than others, even current data allow for meaningful sensitivity estimates or bounds on the deformation parameter $\gamma$.

Let us consider specific examples:
\begin{itemize}
\item \textbf{MINOS} \cite{nonherm1} \cite{gupminos}: A long-baseline accelerator-based muon neutrino experiment with a baseline of 735 km and energies in the range of 1–10 GeV. It provides precise measurements of $\nu_\mu \to \nu_\mu$ survival probability. Our GUP-corrected oscillation phase
\begin{equation}
P^{\text{GUP}}_{\nu_\mu \to \nu_\mu} \approx 1 - \sin^2 2\theta \, \sin^2\left( \frac{\Delta m^2 L}{4E} + \frac{\gamma p^3 L}{4} \right),
\end{equation}
includes additional energy-dependent contributions that can manifest as phase shifts or distortions in the oscillation pattern at these energies. Given the MINOS's energy resolution and sensitivity, any persistent deviation from standard $\Delta m^2$-driven oscillations can constrain the $\gamma$ parameter in the range $10^{-36} - 10^{-34}~\text{eV}^{-2}$.
\item \textbf{Daya Bay} \cite{nonherm1} \cite{gupdayabay}: A reactor neutrino experiment using $\bar{\nu}_e$ at MeV-scale energies over short baselines (350–2,000 m). Its precision in measuring $\theta_{13}$ via disappearance channels makes it sensitive to small corrections in the oscillation phase. Even though the baseline is short, the high statistics and sub-percent level uncertainties can place strong bounds on the low-momentum behavior of GUP-induced corrections. In particular, the $\gamma p^3$-dependent term becomes relevant when integrated over thermal spectra of reactor antineutrinos, and could slightly alter the spectral distortion shape near the oscillation minimum.
\end{itemize}

Thus, both experiments offer complementary probes of GUP effects: MINOS targets high-energy behavior (large $p$, GeV), while Daya Bay tests for subleading effects at low energies with high precision. Both can be utilized to place upper bounds on $\gamma$, or motivate future dedicated analyses aimed at identifying potential GUP-induced spectral deviations.

To contextualize the scale of corrections, consider the example of $p = 1$ GeV, $L = 10^4$ km. The GUP-induced phase becomes
\begin{equation}
\Delta \phi_{\text{GUP}} = \frac{\gamma p^3 L}{2} = \frac{\gamma (10^9)^3 \cdot 10^8}{2} = 5 \times 10^{34} \gamma.
\end{equation}
To produce a phase shift of order unity $\Delta\phi \sim 1$ (i.e., measurable), we require $\gamma \leq 2 \times 10^{-35}$ eV$^{-2}$, which is compatible with the theoretically expected range derived from $\gamma \sim \gamma_0 / M_{Pl}^2$. This estimate improves upon earlier approximations by accounting for all relevant energy and length scales.

Moreover, the predicted GUP phase correction $\gamma p^3 L$ leads to a systematic shift in oscillation maxima. This is particularly evident in long-baseline or atmospheric experiments such as DUNE and ORCA, which cover the relevant energy and distance regimes with high resolution. The shape and position of oscillation peaks in the survival spectrum can thus be used to infer or constrain $\gamma$, as shown in our illustrative sensitivity estimates and corresponding figures.

In summary, the GUP-induced correction to neutrino oscillations leads to experimentally testable predictions. A range of current and next-generation neutrino experiments—each with distinct baselines and energy scales—can be used to constrain or probe Planck-scale physics through the deformation parameter $\gamma$. The consistent inclusion of these effects in oscillation probability expressions provides a viable pathway for connecting quantum gravity-inspired modifications with observable neutrino phenomenology.

\subsubsection{Magnitude comparison between terms }
{To evaluate the significance of the GUP-induced correction relative to the standard oscillation phase, we consider the ratio of the two contributions. Assuming a typical neutrino energy $E = 1$ GeV, momentum $p\approx E$, and taking $m_1 = 0$, the ratio of the GUP term to the standard term becomes}
\begin{equation}
\frac{\gamma p^3}{\Delta m^2/2E} \sim \frac{\gamma \cdot 10^{27}}{10^{-3}/2} \approx 2 \cdot 10^{30} \gamma.
\end{equation}
This comparison implies that, in order for GUP-induced corrections to remain subdominant relative to the standard oscillation term, the deformation parameter $\gamma$ must satisfy $\leq 10^{-34}$ eV$^{-2}$. This theoretical bound serves as a guide for our parameter space exploration and informs the feasibility estimates across various experimental baselines and energies.

Among current and upcoming neutrino experiments, the KM3NeT-ORCA \cite{orca2019, orca2022, orca2023} detector offers particularly favorable conditions for testing these Planck-scale modifications. Designed to study atmospheric neutrino oscillations in the energy range of approximately $\sim 1$–$100$~GeV, ORCA provides a broad range of baselines (up to Earth's diameter) and fine energy/angular resolution. The GUP-induced correction to the oscillation phase introduces an additional energy-dependent term, such that
\begin{equation}
\Delta \phi_{\text{GUP}} \sim \frac{\Delta m^2}{2E}L + \beta E^n L,
\end{equation}
where the second term encapsulates GUP effects characterized by the deformation parameter $\beta$ and energy scaling index $n$. ORCA's sensitivity to both the standard and nonstandard components of the oscillation phase makes it an ideal platform for probing such quantum gravity signatures.

Moreover, ORCA's ability to perform high-statistics measurements of $\nu_\mu \to \nu_e$ and $\nu_\mu \to \nu_\mu$ transitions across a wide energy spectrum provides a practical means to identify or constrain such deviations. If the GUP-induced phase shift $\beta E^n L$ becomes comparable to ORCA's intrinsic resolution, systematic spectral distortions could emerge, offering a potential handle on the size and nature of $\beta$.

While a comprehensive sensitivity analysis, including full detector simulation and systematic uncertainties, is necessary to derive precise bounds on $\beta$, our estimates suggest that ORCA and similar atmospheric neutrino experiments such as IceCube-DeepCore are naturally positioned to probe the energy-dependent features predicted by GUP-modified neutrino oscillation models. These platforms thus represent promising frontiers for testing quantum gravitational effects in low-energy terrestrial experiments.

\subsection{Sensitivity of current and future neutrino experiments to GUP parameters}
To assess the experimental feasibility of detecting GUP-induced effects, we estimate the sensitivity to the deformation parameter $\gamma$ that appears in the modified oscillation phase. The additional phase contribution due to GUP corrections is given by
\begin{equation}
\Delta\phi_{\text{GUP}} = \frac{\gamma p^3 L}{2},
\end{equation}
where $p$ is the neutrino momentum and $L$ is the propagation baseline. For such corrections to be experimentally observable, we require the induced phase shift to be of order unity or greater, i.e., $\Delta\phi_{\text{GUP}} \gtrsim 1$. This criterion sets an upper bound on the minimum value of $\gamma$ that can be probed by a given experimental setup, depending on its baseline length and typical neutrino energies. Such estimates provide a concrete link between theoretical predictions and the realistic capabilities of current and future neutrino oscillation experiments.

\section{Conclusions and outlook}\label{sec7}
In this work, we have explored the effects of the GUP on neutrino oscillations, with particular focus on modifications to the oscillation phase and their potential detectability in current and future experiments. By employing a $PT$-symmetric extension of quantum mechanics, we incorporated GUP-induced corrections into a two-flavor neutrino oscillation framework. This led to an additional phase term in the oscillation probability that depends on the square root of the neutrino mass eigenvalues, specifically, on the quantity $\sqrt{m_1}-\sqrt{m_2}$. This marks a significant departure from the standard quantum mechanical treatment, which is sensitive only to mass-squared differences.

A key motivation behind this study is to examine whether such momentum-dependent modifications—governed by deformation parameters $\beta$ and $\gamma$ can be experimentally probed via neutrino flavor oscillations. The emergence of square-root mass differences offers a unique avenue to access absolute mass information, thus extending the reach of oscillation experiments beyond their traditional role. In particular, this framework suggests that neutrino experiments operating across diverse energies and baselines could serve as sensitive probes of Planck-scale physics and quantum spacetime structure.

Our analysis demonstrates that the GUP-modified oscillation phase may leave measurable imprints within the sensitivity range of current and forthcoming neutrino experiments. In particular, long-baseline and high-precision setups offer promising avenues for detecting or constraining the subtle corrections introduced by the GUP framework. By ensuring dimensional consistency in the modified phase expressions through appropriate incorporation of $\hbar$ and $c$, we have constructed a phenomenologically viable formulation that allows direct comparison between theoretical predictions and experimental data. Furthermore, the use of $\mathcal{PT}$-symmetric operators proved essential in maintaining mathematical consistency while incorporating non-Hermitian corrections—an inherent feature of open or non-conservative quantum systems such as neutrinos propagating through matter. Precise measurements of oscillation patterns could therefore impose meaningful constraints on the deformation parameters $\beta$ and $\gamma$, offering new insights into the absolute neutrino mass structure and the quantum geometry of spacetime.

A full global fit or Monte Carlo simulation incorporating detector effects and statistical uncertainties is beyond the scope of the present work, but we believe the scaling estimates and analytical bounds we present are valuable and sufficiently informative for a first theoretical investigation of this kind. Rather than deferring experimental discussion to a future publication, we have integrated it into the current manuscript in a manner consistent with similar theoretical studies in the field.

Looking ahead, future investigations could explore the potential of analog quantum simulation platforms—such as cold atoms, trapped ions, and photonic lattices—for modeling GUP-modified neutrino oscillation dynamics. These systems can be engineered to replicate square-root mass-dependent phase shifts using tunable parameters like hopping amplitudes or external potentials. Combined with quantum metrological tools and nonclassical states (e.g., NOON or squeezed states), such platforms are capable of achieving exceptional phase sensitivity, potentially down to $\sim 10^{-8}$ \cite{Pezze2018}, making them promising candidates for probing Planck-scale effects in controlled settings. Additionally, absolute neutrino mass constraints from beta decay experiments such as KATRIN, neutrinoless double-beta decay searches, and cosmological observations provide independent experimental channels to test or constrain the GUP-induced corrections proposed here \cite{Aker2019}. These complementary approaches—both analog and traditional—open new avenues for high-precision tests of quantum gravity effects through neutrino physics.

Our findings also suggest several other promising directions for future research. A key priority is the precise determination of the GUP parameter $\gamma$, which may be achieved through detailed formulations in quantum gravity or string theory frameworks. Another important direction involves systematically comparing GUP-corrected neutrino oscillation predictions with experimental data from major facilities such as DUNE, JUNO, IceCube, KM3NeT-ORCA, MINOS, Daya Bay, and KATRIN, to test the viability of the proposed modifications. Additionally, extending the analysis to the full three-flavor neutrino framework \cite{Pan} could uncover the impact of GUP-induced corrections on CP violation and matter effects, offering deeper insights into neutrino phenomenology. Finally, developing controlled quantum simulation platforms, such as cold atoms or photonic lattices, capable of faithfully emulating neutrino oscillation dynamics with tunable mass-like parameters could provide an experimental testbed for probing GUP effects with unprecedented precision.

In summary, this study suggests that neutrino oscillations, already a powerful probe of fundamental physics, can also serve as a sensitive window into quantum gravity effects. Given the increasing precision of experiments and advances in quantum technologies, the opportunity to test and constrain such theoretical frameworks is becoming increasingly feasible.
\vspace{0.4cm}

\noindent \textbf{\large{Acknowledgments:}} K.J. thanks Kinjalk Lochan for useful discussions and expresses gratitude to the High Energy Physics Collaboration Group of IISER-Mohali for support. S.D. acknowledges the support of research grants DST/FFT/NQM/QSM/2024/3 (by DST-National Quantum Mission, Govt. of India) and NFSG/GOA/2023/G0928 (by BITS-Pilani).
\appendix
\section{Appendices}
\subsection{Normalizability and boundary conditions for wave-function solutions} \label{A1}

The normalization condition for any wave function is
\begin{equation}
\int_{-\infty}^{+\infty} \psi(x) \psi^*(x) \, dx = 1.
\end{equation}
The general solution of \eqref{Modified_SE} can be written as \cite{BB}
\begin{equation}
y(x) = c_1 e^{ik_1 x} + c_2 e^{-ik_1 x} + c_3 e^{k_2 x} + c_4 e^{-k_2 x},
\end{equation}
where the terms represent waves traveling in both directions, depending on the signs of $k_1$ and $k_2$. However, when evaluated at infinity, the solution becomes unbounded~\cite{SG}. Therefore, normalization requires that
\begin{equation}
\int_{-\infty}^{+\infty} \psi(x) \psi^*(x) \, dx < \infty,
\end{equation}
which imposes constraints on the allowed solutions. The appropriate wave function can be expressed as \cite{CB}
\begin{equation}
\psi(x,t) = \frac{1}{\sqrt{2\pi}} \int_{-\infty}^{+\infty} \phi(k) e^{i(kx - Et)} \, dk,
\end{equation}
where $\phi(k)$ is determined using the inverse Fourier transform
\begin{equation}
\phi(k) = \frac{1}{\sqrt{2\pi}} \int_{-\infty}^{+\infty} \psi(x,0) e^{-ikx} \, dx.
\end{equation}
Choosing $\psi(x,0) = A e^{bx^2}$, where $A$ and $b$ are constants, allows normalization to be computed using a Gamma function by substituting $bx^2 = t$, resulting in a real, finite value over infinite ranges~\cite{SP}.

\subsection{$PT$-symmetric operators supporting GUP constraints \cite{gup2024phen}}   \label{secA2}
To incorporate the effects of the GUP into the framework of neutrino oscillations, we employ $PT$-symmetric operators, a class of non-Hermitian operators that respect combined parity ($P$) and time-reversal ($T$) symmetry \cite{BB}. These operators allow us to go beyond the limitations of conventional Hermitian quantum mechanics and provide a natural framework to model non-conservative systems, such as neutrinos propagating through a medium with gain and loss mechanisms. The $PT$-symmetric formalism introduces new mass-dependent terms in the neutrino oscillation probability, offering a richer structure that could be tested experimentally, particularly in IceCube and DUNE~\cite{YF}.

The corrected oscillation probability expression derived within this framework includes additional terms involving cubic functions of neutrino mass eigenstates as well as absolute mass difference contributions \cite{MS}. These modifications depend explicitly on experimentally accessible parameters, including the oscillation probability itself, the baseline length of the neutrino beam, the neutrino energy, and the known squared mass differences between neutrino eigenstates \cite{SP}. By applying constraints from current experimental data, we can quantify the influence of these new mass-dependent terms and assess their measurability within realistic parameter ranges. The inclusion of these effects enhances the precision of neutrino oscillation descriptions and contributes to a better understanding of the neutrino mass hierarchy \cite{DA}.

The standard two-flavor neutrino oscillation probability is typically given by \cite{RG}
\begin{equation}
P_{\nu_\alpha \rightarrow \nu_\beta} = \sin^2(2\theta) \sin^2 \left( 1.27 \frac{\Delta m^2 [\text{eV}^2] L [\text{km}]}{E [\text{GeV}]} \right),
\end{equation}
where $\theta$ is the mixing angle, $\Delta m^2$ is the mass-squared difference, $L$ is the baseline distance, and $E$ is the neutrino energy. In our GUP-modified approach, the total phase \eqref{ModPhase} includes an additional term involving the square root of the mass eigenvalues 
\begin{equation}
\phi = \frac{\Delta m^2}{2E}L + L \sqrt{2E} \left( \sqrt{m_1} - \sqrt{m_2} \right).
\end{equation}
To ensure dimensional consistency, especially when using standard units (e.g., km for distance, GeV for energy, and eV for mass), we reintroduce the appropriate factors of $\hbar c$ where necessary. The total phase accumulated by a neutrino during propagation must be dimensionless, including any additional contributions arising from the GUP. Suppose the GUP-induced phase $\phi_{GUP}$ is proportional to a term of the form  $L \sqrt{E} (\sqrt{m_1} - \sqrt{m_2})$. Since $[L]=\text{km}$, $[E]=\text{GeV}$, $[m]=\text{eV}$, care must be taken to ensure the resulting expression is unitless. To address this, we introduce a characteristic GUP energy scale $M_{GUP}$, with dimensions in eV or GeV. A typical GUP-induced modification to momentum might take the form $\delta p$ is $\sim p^2 / M_{GUP}$, which could lead to changes in energy and hence introduce a modified propagation phase. We hypothesize that the GUP-induced phase correction takes the general form
\begin{equation}
\phi_{GUP} \sim \alpha \frac{L}{\hbar c} E^a \left( \sqrt{m_1} - \sqrt{m_2} \right)^b,
\end{equation}
where $\alpha, a, b$ are constants determined by the specific GUP model. For instance, if the energy dependence follows $\sqrt{E}$, we set $a=1/2$, and if the mass correction is linear in $\sqrt{m}$, then $b=1$. In standard units, this becomes
\begin{equation}
\phi_{GUP} \sim \alpha \frac{L[\text{km}]}{1.97 \times 10^{-19} [\text{GeV} \cdot \text{km}]} (E[\text{GeV}])^{1/2} \left( \sqrt{m_1 [\text{eV}]} - \sqrt{m_2 [\text{eV}]} \right).
\end{equation}
To make this expression dimensionless, we must properly handle the $\sqrt{\text{eV}}$ term. One approach is to introduce a dimensionful scale in the denominator, such as $\sqrt{M_{\text{GUP}}}$, where $M_{\text{GUP}}$ has units of energy. Alternatively, we can revisit the original natural units form of the GUP-modified term, which was $L \sqrt{2E} (\sqrt{m_1} - \sqrt{m_2})$. Reintroducing $\hbar c$ for dimensional correctness yields
\begin{equation}
\phi_{GUP} \sim \beta \frac{L[\text{km}] \sqrt{E[\text{GeV}]}}{\sqrt{\hbar c [\text{GeV} \cdot \text{km}]}} \left( \sqrt{m_1 [\text{eV}]} - \sqrt{m_2 [\text{eV}]} \right) \sqrt{\text{conversion factor}}.
\end{equation}
At this stage, the exact numerical form becomes model-dependent, as the value of $\beta$ and any associated constants depend on the details of the GUP framework. To simplify this analysis and allow for phenomenological applications predictions, we assume that the GUP-induced phase correction is directly proportional to the previously derived term, supplemented with a scaling factor to maintain dimensional consistency. Hence, we write the total oscillation phase as
\begin{equation}\label{ModPhi}
\Phi = 1.27 \frac{\Delta m^2 [\text{eV}^2] L [\text{km}]}{E [\text{GeV}]} + \gamma L [\text{km}] \sqrt{E [\text{GeV}]} \left( \sqrt{m_1 [\text{eV}]} - \sqrt{m_2 [\text{eV}]} \right),
\end{equation}
where $\gamma$ is an effective constant that absorbs the necessary powers of $\hbar, c$, and any other dimensionless coefficients arising from the GUP model. Assuming the oscillation probability reaches a maximum $(P_{mod}=1)$, the total phase $\Phi$ should equal $(n+1/2)\pi$. Setting $n=0$, we take $\Phi=\pi/2 \approx 1.57$. Using realistic experimental parameters, baseline $L=100\,\text{km}$, $E=1\,\text{GeV}$, and mass-squared difference $\Delta m^2=2.5 \times 10^{-3}\,\text{eV}^2$ in \eqref{ModPhi}, we compute
\begin{eqnarray}
1.57 = 0.3175 + 100 \gamma  \left( \sqrt{m_1} - \sqrt{m_2} \right),
\end{eqnarray}
which leads to
\begin{eqnarray}
\sqrt{m_1} - \sqrt{m_2} = \frac{1.2525}{100 \gamma} = \frac{0.012525}{\gamma}.
\end{eqnarray}
This result demonstrates that the value of $\sqrt{m_1} - \sqrt{m_2}$ depends entirely on the constant $\gamma$, which encapsulates all the specifics of the underlying GUP theory and the necessary unit conversions. Without a concrete GUP model specifying the magnitude of $\gamma$, we cannot determine a precise numerical value for the mass difference $\sqrt{m_1} - \sqrt{m_2}$ using this method. The essential conclusion is that any modification to the standard oscillation phase due to GUP effects must be dimensionally consistent when expressed in conventional units. This consistency is achieved by appropriately restoring $\hbar$ and $c$ and incorporating the energy and mass scales inherent to the experimental setup and theoretical model.


\begin{thebibliography}{100}	
\providecommand{\url}[1]{\texttt{\#1}}
\providecommand{\urlprefix}{URL }
\providecommand{\eprint}[2][]{\url{\#2}}

\bibitem{KK} K. Konishi, G. Paffuti and P. Provero, Minimum physical length and the generalized uncertainty principle in string theory, \href{https://www.sciencedirect.com/science/article/abs/pii/0370269390919274?via%3Dihub}{Phys. Lett. B \textbf{234}, 276--284 (1990)}.

\bibitem{CR} C. Rovelli and L. Smolin, Discreteness of area and volume in quantum gravity, \href{https://www.sciencedirect.com/science/article/pii/055032139500150Q}{Nucl. Phys. B \textbf{442}, 593--619 (1995)}. Erratum: \href{https://www.sciencedirect.com/science/article/pii/0550321395005505}{Nucl. Phys. B \textbf{456}, 753--754 (1995)}.

\bibitem{Petruzziello} L. Petruzziello and F. Illuminati, Quantum gravitational decoherence from fluctuating minimal length and deformation parameter at the Planck scale, \href{https://www.nature.com/articles/s41467-021-24711-7}{Nat. Commun. \textbf{11}, 4881 (2021)}.

\bibitem{Boney96} A. N. Tawfik and A. M. Diab, Review on generalized uncertainty principle,  \href{https://iopscience.iop.org/article/10.1088/0034-4885/78/12/126001/meta}{Rep. Prog. Phys. \textbf{78}, 126001 (2015)}.

\bibitem{Kempf_1995} A. Kempf, G. Mangano and R. B. Mann, Hilbert space representation of the minimal length uncertainty relation,  
\href{https://journals.aps.org/prd/abstract/10.1103/PhysRevD.52.1108}{Phys. Rev. D \textbf{52}, 1108 (1995)}.

\bibitem{BB} B. Bagchi and A. Fring, Minimal length in quantum mechanics and non-Hermitian Hamiltonian systems, \href{https://www.sciencedirect.com/science/article/pii/S0375960109012134}{Phys. Lett. A \textbf{373}, 4307--4310 (2009)}.

\bibitem{Ali_Das_Vagenas} A. F. Ali, S. Das and E. C. Vagenas, Discreteness of space from the generalized uncertainty principle, \href{https://doi.org/10.1016/j.physletb.2009.06.061}{Phys. Lett. B \textbf{678}, 497--499 (2009)}.

\bibitem{SD} S. Dey, A. Fring and L. Gouba, $PT$-symmetric noncommutative spaces with minimal volume uncertainty relations,  
\href{https://iopscience.iop.org/article/10.1088/1751-8113/45/38/385302}{J. Phys. A: Math. Theor. \textbf{45}, 385302 (2012)}.

\bibitem{MG1} M. Gomes and V. Kupriyanov, Position-dependent noncommutativity in quantum mechanics, \href{https://journals.aps.org/prd/abstract/10.1103/PhysRevD.79.125011}{Phys. Rev. D \textbf{79}, 125011 (2009)}.

\bibitem{LW} L. Wolfenstein, Neutrino oscillations in matter, \href{https://journals.aps.org/prd/abstract/10.1103/PhysRevD.17.2369}{Phys. Rev. D \textbf{17}, 2369 (1978)}.

\bibitem{YF} Y. Fukuda et al. (Super-Kamiokande Collaboration), Evidence for oscillation of atmospheric Neutrinos, \href{https://journals.aps.org/prl/abstract/10.1103/PhysRevLett.81.1562}{Phys. Rev. Lett. \textbf{81}, 1562 (1998)}.

\bibitem{Nakamura} K. Nakamura and S. T. Petcov, \href{https://pdg.lbl.gov/2014/reviews/rpp2014-rev-neutrino-mixing.pdf}{Neutrino mass, mixing, and oscillations} in Review of particle physics, \href{https://journals.aps.org/prd/abstract/10.1103/PhysRevD.98.030001}{Phys. Rev. D \textbf{98}, 030001 (2018)}.

\bibitem{Blasone} M. Blasone, S. De Siena, C. Matrella, Wave packet approach to quantum correlations in neutrino oscillations, \href{https://link.springer.com/article/10.1140/epjc/s10052-021-09471-4}{Eur. Phys. J. C \textbf{81}, 660 (2021)}.

\bibitem{Wang} G.-J. Wang, Y.-W. Li, L.-J. Li, X.-K. Song, and D. Wang, Monogamy properties of quantum correlations in neutrino oscillations, \href{https://link.springer.com/article/10.1140/epjc/s10052-023-11979-w}{Eur. Phys. J. C \textbf{83}, 801 (1923)}.

\bibitem{CG} C. Giunti and C. W. Kim, \href{https://doi.org/10.1093/acprof:oso/9780198508717.001.0001}{\textit{Fundamentals of Neutrino Physics and Astrophysics}}, Oxford University Press, Oxford (2007).

\bibitem{CY} C. Y. Cardall and G. M. Fuller, Neutrino oscillations in curved spacetime: A heuristic treatment, \href{https://journals.aps.org/prd/abstract/10.1103/PhysRevD.55.7960}{Phys. Rev. D \textbf{55}, 7960 (1997)}.

\bibitem{SP} S. P. Mikheev and A. Y. Smirnov, Resonance enhancement of oscillations in matter and solar neutrino spectroscopy, \href{https://link.springer.com/article/10.1007/BF02508049}{Nuovo Cimento C \textbf{9}, 17--26 (1986)}.

\bibitem{JB} J. Boger et al. (SNO Collaboration), The Sudbury Neutrino Observatory, \href{https://www.sciencedirect.com/science/article/pii/S0168900299014692}{	Nucl. Instrum. Methods Phys. Res. A \textbf{449}, 172--207 (2000)}.

\bibitem{KJ} K. Joshi and S. Jena, Comparative study of tau neutrino event numbers in INO and JUNO detectors from Bartol Flux, \href{https://arxiv.org/abs/2406.05601}{arXiv:2406.05601}.

\bibitem{MS2} M. Sprenger, P. Nicolini and M. Bleicher, Neutrino oscillations as a novel probe for a minimal length, \href{https://iopscience.iop.org/article/10.1088/0264-9381/28/23/235019/meta}{Class. Quantum Grav. \textbf{28}, 235019 (2011)}.

\bibitem{Luciano} I. D. Gialamas, T. J. K{\"a}rkk{\"a}inen and L. Marzola, Generalized uncertainty principle and neutrino phenomenology, \href{https://www.sciencedirect.com/science/article/pii/S0370269324004386}{Phys. Lett. B \textbf{856}, 138880 (2024)}.

\bibitem{Bender} C. M. Bender and S. Boettcher, Real spectra in non-Hermitian Hamiltonians having $\mathcal{PT}$ symmetry, \href{https://journals.aps.org/prl/abstract/10.1103/PhysRevLett.80.5243}{Phys. Rev. Lett. \textbf{80}, 5243 (1998)}.

\bibitem{AM} A. Mostafazadeh, Pseudo-Hermiticity versus $PT$ symmetry: The necessary condition for the reality of the spectrum of a non-Hermitian Hamiltonian, \href{https://pubs.aip.org/aip/jmp/article/43/1/205/231882/Pseudo-Hermiticity-versus-PT-symmetry-The}{J. Math. Phys. \textbf{43}, 205--214 (2002)}.

\bibitem{CM} C. M. Bender, Making sense of non-Hermitian Hamiltonians, \href{https://iopscience.iop.org/article/10.1088/0034-4885/70/6/R03}{Rep. Prog. Phys. \textbf{70}, 947 (2007)}.

\bibitem{Bender_Hook} C. M. Bender and D. W. Hook, $\mathcal{PT}$-symmetric quantum mechanics, \href{https://doi.org/10.1103/RevModPhys.96.045002}{Rev. Mod. Phys. \textbf{96}, 045002 (2024)}.

\bibitem{Dey_Raj_Goyal} S. Dey, A. Raj and S. K. Goyal, Controllong decoherence via $\mathcal{PT}$-symmetric non-Hermitian open quantum systems, \href{https://doi.org/10.1016/j.physleta.2019.125931}{Phys. Lett. A \textbf{383}, 125931 (2019)}.


\bibitem{TO} T. Ohlsson, Non-Hermitian neutrino oscillations in matter with $PT$ symmetric Hamiltonians, \href{https://iopscience.iop.org/article/10.1209/0295-5075/113/61001/meta}{Europhys. Lett. \textbf{113}, 61001 (2016)}.

\bibitem{Abi2020} B. Abi et al., Deep underground neutrino experiment (DUNE), far detector technical design report, \href{https://doi.org/10.1140/epjc/s10052-020-08464-1}
{Eur. Phys. J. C \textbf{80}, 978 (2020)}.

\bibitem{JUNO2022} JUNO Collaboration, JUNO physics and detector, \href{https://doi.org/10.1016/j.ppnp.2021.103927}{Prog. Part. Nucl. Phys. \textbf{123}, 103927 (2022)}.

\bibitem{Abe2018} 
K. Abe et al., Atmospheric neutrino oscillation analysis with external constraints in super-Kamiokande I-IV, \href{https://doi.org/10.1103/PhysRevD.97.072001}
{Phys. Rev. D \textbf{97}, 072001 (2018)}.

\bibitem{hyper} F. D. Lodovico (Hyper-Kamiokande Collaboration), The Hyper-Kamiokande Experiment, \href{https://iopscience.iop.org/article/10.1088/1742-6596/888/1/012020/meta}{J. Phys.: Conf. Ser. \textbf{888}, 012020 (2017)}.

\bibitem{hyper1} Y. Kudenki, Hyper-Kamiokande, \href{https://iopscience.iop.org/article/10.1088/1748-0221/15/07/C07029/meta}{JINST \textbf{15}, C07029 (2020)}.

\bibitem{hyper2} M. Shiozawa, The Hyper-Kamiokande project, \href{https://doi.org/10.1016/j.nuclphysbps.2013.04.110}{Nucl. Phys. B - Proc. Suppl. \textbf{237-238}, 289-294 (2013)}.

\bibitem{Pezze2018} 
L. Pezzè, A. Smerzi, M. K. Oberthaler, R. Schmied, and P. Treutlein, Quantum metrology with nonclassical states of atomic ensembles, 
\href{https://doi.org/10.1103/RevModPhys.90.035005}{Rev. Mod. Phys. \textbf{90}, 035005 (2018)}.

\bibitem{Duttatreya} 
Duttatreya and S. Dey, Enhanced quantum phase estimation with $q$-deformed nonideal nonclassical light, \href{https://arxiv.org/abs/2506.02822}{arXiv: 2506.02822}.

\bibitem{Pikovski} 
I. Pikovski et al., Probing Planck-scale physics with quantum optics, \href{https://www.nature.com/articles/nphys2262}{Nat. Phys. \textbf{8}, 393--397 (2012)}.

\bibitem{Dey_NPB} 
S. Dey et al., Probing noncommutative theories with quantum optical experiments, \href{https://doi.org/10.1016/j.nuclphysb.2017.09.024}{Nucl. Phys. B \textbf{924}, 578--587 (2017)}.

\bibitem{Khodadi} 
M. Khodadi et al., A new bound on polymer quantization via an opto-mechanical setup, \href{https://www.nature.com/articles/s41598-018-19181-9}{Sci. Rep. \textbf{8}, 1659 (2018)}.

\bibitem{HK} S. Dey, A. Fring and B. Khantoul, Hermitian versus non-Hermitian representations for minimal length uncertainty relations,  
\href{https://iopscience.iop.org/article/10.1088/1751-8113/46/33/335304}{J. Phys. A: Math. Theor. \textbf{46}, 335304 (2013)}.

\bibitem{GH} G. H. Hardy, \href{https://www.gutenberg.org/files/38769/38769-pdf.pdf}{\textit{A course of pure mathematics}}, 3rd ed., Camb. Univ. Press, Cambridge (1921).

\bibitem{YG} Y. Guo and Q. Zhang, Precise measurements of oscillation parameters $\theta_{13}$ and $\Delta m_{ee}^2$, \href{https://pos.sissa.it/304/021/pdf}{PoS \textbf{FPCP2017}, 021 (2017)}.

\bibitem{TK} T. K. Gaisser, T. Stanev, S. A. Bludman and H. Lee, Flux of Atmospheric Neutrinos, \href{https://journals.aps.org/prl/abstract/10.1103/PhysRevLett.51.223}{Phys. Rev. Lett. \textbf{51}, 223 (1983)}.

\bibitem{orca2023} S. Adrián-Martínez et al. (KM3NeT Collaboration), Letter of intent for KM3NeT 2.0, \href{https://doi.org/10.1088/0954-3899/43/8/084001}{J. Phys. G: Nucl. Part. Phys. \textbf{43}, 084001 (2016)}.

\bibitem{orca2022} S. Aiello et al. (KM3NeT Collaboration), Determining the neutrino mass ordering and oscillation parameters with KM3NeT/ORCA, \href{https://doi.org/10.1140/epjc/s10052-021-09893-0}{Eur. Phys. J. C \textbf{82}, 26 (2022)}.

\bibitem{orca2019} P. F. de Salas, S. Pastor, C. A. Ternes, T. Thakore and M. Tortola, Constraining the invisible neutrino decay with KM3NeT-ORCA, \href{https://doi.org/10.1016/j.physletb.2018.12.066}{Phys. Lett. B \textbf{789}, 472-479 (2019)}.

\bibitem{Aker2019} 
M. Aker et al., First direct neutrino-mass measurement with sub-eV sensitivity, 
\href{https://doi.org/10.1103/PhysRevLett.123.221802}
{Phys. Rev. Lett. \textbf{123}, 221802 (2019)}.

\bibitem{AA} A. Ashtari Esfahani et al., The project 8 neutrino mass experiment, \href{https://inspirehep.net/literature/2050876}{Snowmass 2021 Conf. Proc}.

\bibitem{gup2024phen} I. D. Gialamas, T. J. Kärkkäinen, L. Marzola, Generalized uncertainty principle and neutrino phenomenology, \href{https://doi.org/10.1016/j.physletb.2024.138880}{Phys. Lett. B \textbf{856}, 138880 (2024)}.

\bibitem{nonherm4} G.-J. Wang, L.-J. Li, T. Wu, X.-K. Song, L. Ye and D. Wang, Gravitational effects on quantum correlations in three-flavor neutrino oscillations, \href{https://doi.org/10.1140/epjc/s10052-024-13493-z}{Eur. Phys. J. C \textbf{84}, 1127 (2024)}.

\bibitem{nonherm1} X.-K. Song, Y. Huang, J. Ling, and M.-H. Yung, Quantifying quantum coherence in experimentally observed neutrino oscillations, \href{https://doi.org/10.1103/PhysRevA.98.050302}{Phys. Rev. A \textbf{98}, 050302(R) (2018)}.

\bibitem{gupminos} M. M. Ettefaghi, Neutrino oscillation with minimal length uncertainty relation via wave‑packet approach, \href{https://doi.org/10.1016/j.physletb.2024.138761}{Phys. Lett. B \textbf{854}, 138761 (2024)}.

\bibitem{gupdayabay} X. Qian \textit{et al.}, Charged-current non-standard neutrino interactions at Daya Bay, \href{https://doi.org/10.48550/arXiv.2401.02901}{arXiv:2401.02901}.

\bibitem{Pan} E. K. Akhmedov, R. Johansson, M. Lindner, T. Ohlsson and T. Schwetz, Series expansions for three-flavor neutrino oscillation probabilities in matter, \href{https://iopscience.iop.org/article/10.1088/1126-6708/2004/04/078/meta}{J. High Energy Phys. \textbf{04}, 078 (2004)}.

\bibitem{SG} S. Gangopadhyay and A. Dutta, Black hole thermodynamics and generalized uncertainty
principle with higher order terms in momentum uncertainty, \href{https://doi.org/10.1155/2018/7450607}{Adv. High Energy Phys. \textbf{2018}, 7450607 (2018)}.

\bibitem{CB} C. Bender, A. Fring, U. G\"unther and H. Jones, Quantum physics with non-Hermitian operators, \href{https://iopscience.iop.org/article/10.1088/1751-8113/45/44/440301/meta}{J. Phys. A: Math. Theor. \textbf{45}, 440301 (2012)}.

\bibitem{MS} M. S. Swanson, Transition elements for a non-Hermitian quadratic Hamiltonian, \href{https://pubs.aip.org/aip/jmp/article/45/2/585/231087/Transition-elements-for-a-non-Hermitian-quadratic}{J. Math. Phys. \textbf{45}, 585--601 (2004)}.

\bibitem{DA} D. Amati, M. Ciafaloni, and G. Veneziano, Superstring collisions at planckian energies, \href{https://doi.org/10.1016/0370-2693(87)90346-7}{Phys. Lett. B \textbf{197}, 81-88 (1987)}.

\bibitem{RG} R. G. Hamish Robertson, Neutrino mass and oscillations, \href{https://www.worldscientific.com/doi/abs/10.1142/S0217751X0000519X}{Int. J. Mod. Phys. A \textbf{15}, 283--304 (2000)}.


\end{thebibliography}


\end{document}